\documentclass[aps,prl,twocolumn,showpacs,superscriptaddress]{revtex4}
\pdfoutput=1
\usepackage{graphicx}
\newcommand{\kv}{\mathbf{k}}
\usepackage{amsmath} \newcommand{\EqLabel}[1]{\label{#1}}

\begin{document}
 
\title{Holstein polaron in the presence of disorder} 

\author{Mona Berciu}

\affiliation{ Department of
Physics and Astronomy, University of British Columbia, Vancouver, BC,
Canada, V6T~1Z1}

\author{Andrei S. Mishchenko}

\affiliation{Cross-Correlated Materials Research Group (CMRG), ASI,
  RIKEN, Wako 351-0198, Japan}
\affiliation{Russian Research Centre ``Kurchatov Institute'', 
123182 Moscow, Russia}

\author{Naoto Nagaosa}

\affiliation{Cross-Correlated Materials Research Group (CMRG), ASI,
  RIKEN, Wako 351-0198, Japan}

\affiliation{Department of Applied Physics, University of Tokyo, 7-3-1
Hongo, Bunkyo-ku, Tokyo 113, Japan}

\date{\today}
 
\begin{abstract}
Non-local, inhomogeneous and retarded response observed in experiments
is reproduced by introducing the Inhomogeneous Momentum Average (IMA) method to study
single polaron problems with disorder in the on-site potential and/or
spatial variations of the electron-phonon couplings and/or  phonon
frequencies.  We show that the electron-phonon coupling gives rise to
an additional inhomogeneous, strongly retarded potential, which makes
 instant approximations questionable.  The accuracy of
IMA is demonstrated by comparison with results from the approximation
free Diagrammatic Monte Carlo (DMC) method. Its simplicity allows for
easy study of many problems that were previously unaccessible. As an
example, we show how inhomogeneities in the electron-phonon coupling
lead to nonlocal, retarded response in scanning tunneling microscopy
(STM) images.
\end{abstract}

\pacs{71.38.-k, 72.10.Di, 63.20.kd}

\maketitle

Understanding the nature of the materials under the focus of current
basic research, as well as the development of novel applications, is
unambiguously linked to the physics of quasiparticles in disordered
systems and coupled to bosonic modes of different origins.  Thus, the
manganites which exhibit colossal magnetoresistance are doped
materials \cite{Dagot01} with considerable electron-phonon (el-ph) as
well as
electron-magnon and electron-orbitron
couplings \cite{Mil96}; the interplay between disorder
\cite{Uehara99} and the coupling to bosons manifests itself in the
peculiarities of their phase diagram \cite{Moto03}.  Similarly, the
underdoped high temperature cuprate superconductors are inhomogeneous
materials \cite{Pan01} with rather strong \cite{Gunna08} and
inhomogeneous~\cite{bis2} coupling to phonons. Another example concerns charge transport in organic
thin-film transistors \cite{Matsu08}, which is dominated by polaron
jumps between different potential traps \cite{Horo91}.  The importance
of the interplay between disorder and coupling to boson fields is
magnified by the fact that weak coupling to a boson mode that is
relatively unimportant in a clean system may result in dramatic
effects in disordered compounds \cite{Mish}.

The tremendous difficulties in treating the problem of a polaron in
the presence of even a single impurity resulted in the 30 year delay
between the first results based on the adiabatic approximation
\cite{Toy79} and the recent approximation free solution by the DMC
method \cite{Mish}.  However, the current level of technology requires
theoretical predictions for numerous systems whose inhomogeneity is
not limited to a single impurity but includes any form of spatial
inhomogeneity, plus potential wells or barriers representing surfaces,
interfaces or quantum dot wells, besides cases where both the energy
of the bosonic mode and its coupling to the electron are inhomogeneous
\cite{Bal06}. Solving such general problems for large systems by
numerical methods is still effectively impossible.

In this Letter we study accurately yet efficiently the single Holstein polaron problem with
disorder in the potential, the strength of the coupling
constant and the frequency of the phonon by 
developing the Inhomogeneous Momentum Average (IMA)
method.  The method is based on the Momentum Average (MA)
approximation used to study translationally invariant systems, like
Holstein \cite{MB1} and more general models \cite{MB5}.  IMA takes any
potential inhomogeneity into account {\it exactly} and can also handle
spatial variations of the coupling constant and of the frequency of
the boson modes.  Comparing results of IMA with approximation free
data from DMC allows us to gauge its accuracy. We perform this
comparison in one-dimension (1D) where the worst accuracy of IMA is
expected \cite{MB1}. The IMA approximation can then be systematically improved
\cite{MB2} so that in combination with DMC \cite{Mish} one gets an
accurate and fast tool to study all the systems described above within
the framework of a controllable and efficient approximation scheme.

Given the low computational cost of IMA, a rapid
scan of large regions of the parameter space is now possible. To
compare with experimental observations, we compute STM images of inhomogeneous systems of large
enough sizes to render DMC studies impractical, due to the enormous
computational cost required for the analytic continuation method
\cite{MPSS}.  This allows us to prove the nonlocal nature of a
system's response to inhomogeneities and demonstrate the importance of
the retardation effects.
        
To avoid cumbersome expressions we consider one impurity in an
otherwise homogeneous 1D system, indicating generalizations  where suitable. The  Hamiltonian is:
\begin{equation}
\EqLabel{e1} {\cal H} ={\cal H}_0 + \hat{U}_0 + \hat{V}_{\rm el-ph}
\end{equation}
where ${\cal H}_0 = -t \sum_{\langle i,j\rangle}
(c^\dagger_ic_j+h.c.)+ \Omega \sum_{i}^{} b_i^\dagger b_i$ is the free
part, $\hat{U}_0 = -Uc_0^\dagger c_0$  the
on-site attraction to the impurity placed at site $i=0$, and $\hat{V}_{\rm
el-ph} = g\sum_{i}^{} c_i^\dagger c_i \left(b_i^\dagger + b_i\right)$
is  the  el-ph interaction. The electron's spin is irrelevant.

The goal is to compute the retarded Green's function:
\begin{equation}
\EqLabel{e5} G(i,j,\omega) = \langle 0 | c_i \hat{G}(\omega)
c_j^\dagger|0\rangle = \sum_{\alpha}^{ } \frac{\langle 0| c_i
|\alpha\rangle\langle\alpha | c_j^\dagger|0\rangle}{ \omega - E_\alpha
+i\eta},
\end{equation}
where $\hat{G}(\omega) = \left[ \omega - {\cal H} +
i\eta\right]^{-1}$ and $\hbar=1$, because it has information on the
single electron eigenstates ${\cal H} 
|\alpha\rangle = E_\alpha |\alpha\rangle$. Also, the local density of states
(LDOS) $A(i,\omega) = -{1\over \pi} {\rm Im} G(i,i,\omega)$ is measured
directly by STM.

We first introduce two additional Green's functions. One is the free
electron Green's function
\begin{equation}
\nonumber G_0(i,j,\omega) = \langle 0 | c_i \hat{G}_0(\omega)
c_j^\dagger|0\rangle = {1\over 2\pi}\int_{-\pi}^{\pi} dk {e^{i{k}
({R}_i-{R}_j)}\over \omega - \epsilon_{{k}} + i\eta}
\end{equation}
where 
$\epsilon_k = - 2t \cos k$ for nearest neighbor hopping. Generalizations to other dispersions and
higher dimensionality are trivial.
The second  is the ``disorder'' Green's function $
 G_{\rm d}(i,j,\omega) = \langle 0 | c_i \hat{G}_{\rm d}(\omega)
 c_j^\dagger|0\rangle$, corresponding to ${\cal H}_{\rm d}={\cal H}|_{g=0}={\cal H}_0 +
 \hat{U}_0$.  Using Dyson's identity $\hat{G}_{\rm
 d}(\omega) = \hat{G}_0(\omega) + \hat{G}_{\rm d}(\omega) \hat{U}_0
 \hat{G}_0(\omega)$ straightforwardly leads to: 
\begin{equation}
\EqLabel{9} G_{\rm d}(i,j,\omega) = G_0(i-j,\omega) -
U\frac{G_0(i,\omega)G_0(j,\omega)}{1+ U G_0(0,\omega)},
\end{equation}
since $G_0(i,j,\omega) = G_0(i-j,\omega) = G_0(j-i,\omega)$ (the
second equality holds if time reversal symmetry is obeyed).  Eq. (\ref{9}) is valid in
any dimension, if the appropriate $G_0$ is used.  For more complicated
disorder potentials one can find $G_{\rm d}(i,j,\omega)$ by a suitable
generalization: ${\cal H}_{\rm d}$ is a quadratic Hamiltonian and can always be
diagonalized.  Thus, $G_{\rm d}$ is known and treats the
on-site disorder {\em exactly}.

Since ${\cal H}= {\cal H}_{\rm d} + \hat{V}_{\rm el-ph}$, we can now
  proceed to calculate the desired Green's function in terms of
  $G_{\rm d}(i,j,\omega)$. Using Dyson's identity once, we find:
\begin{equation}
\EqLabel{i1} G(i,j,\omega) = G_{\rm d}(i,j,\omega) + g\sum_{j_1}^{}
F_1(i,j_1,\omega) G_{\rm d}(j_1,j,\omega)
\end{equation}
where $F_n(i,j,\omega) = \langle 0| c_i\hat{G}(\omega) c_{j
}^\dagger b^{\dagger n}_j |0\rangle$, with $F_0(i,j,\omega)=G(i,j,\omega)$.
Using the Dyson identity again, we find that:
\begin{eqnarray}
\nonumber F_n(i,j,\omega) = g\sum_{j_1\ne j}^{} G_{\rm d}(j_1,j,\omega
- n\Omega) \langle 0|c_i \hat{G}(\omega) c_{j_1}^\dagger
b_{j_1}^\dagger b_j^{\dagger n}|0\rangle && \\ \EqLabel{gg1} + g
G_{\rm d}(j,j,\omega-n\Omega) \left[nF_{n-1}(i,j,\omega) +
  F_{n+1}(i,j,\omega)\right]. \hspace{5mm}&&
\end{eqnarray}
Within the IMA$^{(0)}$
 approximation, we set in all these equations $G_{\rm d}(j_1,j,\omega-n
\Omega)\rightarrow 0$ if $j\ne j_1$ and $n>0$. This is a good low-energy approximation, because the
ground-state (GS) of the polaron is below the spectrum of ${\cal
H}_{\rm d}$ and so for $\omega \sim E_{GS}$, $G_{\rm
d}(i,j,\omega-n\Omega)$ decreases exponentially with the distance
$|i-j|$, the decrease being steeper for larger $n$.  In other words,
IMA$^{(0)}$ ignores exponentially small
terms.   Like MA$^{(0)}$,  IMA$^{(0)}$ is in fact accurate  at all
energies because it obeys multiple sum rules~\cite{MB1,MB2}. It also
 becomes exact both for  $g\rightarrow 0$ and $t\rightarrow 0$.

With this approximation only the  $F_n$ functions survive in
Eq. (\ref{gg1}), whose general solution is then $F_n(i,j,\omega) =
A_n(j,\omega) F_{n-1}(i,j,\omega)$~\cite{MB1}. The continued fractions:
\begin{equation}
\EqLabel{i5} A_n(j,\omega)=\frac{ngG_{\rm
d}(j,j,\omega-n\Omega)}{1-gG_{\rm d}(j,j,\omega-n\Omega)
A_{n+1}(j,\omega) }
\end{equation}
are simple to compute. We now insert $F_1(i,j,\omega) =
A_1(j,\omega) G(i,j,\omega)$ in Eq. (\ref{i1}), resulting in
$G(i,j,\omega) = G_{\rm d}(i,j,\omega) +
g\sum_{j_1}^{}G(i,j_1,\omega) A_1(j_1,\omega) G_{\rm
d}(j_1,j,\omega)$. To make this system of coupled equations
converge fast with the cutoff in
$j_1$, we define
\begin{equation}
\EqLabel{i8} A_n(\omega)=A_n(j,\omega)|_{U=0} =
A_n(j,\omega)|_{|j|\rightarrow \infty},
\end{equation}
since if $U=0$, then $G_{\rm d}(j, j,\omega)= G_0(j,j,\omega) =
G_0(0,\omega)$ irrespective of $j$. The same holds for
finite $U$ but $|j|\rightarrow \infty$, since sites located very far from
the impurity
are not sensitive to its presence at the origin.
Introducing the ``effective interaction'' potential:
\begin{equation}
\EqLabel{ii9} v_0 (j,\omega) =g A_1(j,\omega)-\Sigma_{\rm MA^{(0)}}(\omega)
\end{equation}
where the bulk  MA$^{(0)}$ self-energy is
$\Sigma_{\rm MA^{(0)}}(\omega) = g A_1(\omega)$ \cite{MB2},  we can
rewrite the equation for $G(i,j,\omega)$ as: 
\begin{equation}
\EqLabel{ma0} G(i,j,\omega) = G_{\rm d}(i,j,\tilde{\omega})+
 \sum_{j_1}^{}G(i,j_1,\omega)v_0(j_1,\omega) G_{\rm
 d}(j_1,j,\tilde{\omega})
\end{equation}
where $\tilde{\omega}= \omega-\Sigma_{\rm MA^{(0)}}(\omega)$.  This
equation is very efficient to solve numerically, because
$v_0(j,\omega)\rightarrow 0$ rapidly with increasing $|j|$. In fact, a
cutoff $|j| \le 5$ suffices for convergence, although  a cutoff of 0
or 1 does {\em not}, showing that $v_0(j,\omega)$ is spread over a few
sites 
around the impurity. Note that inhomogeneities in the
values of $g$ and $\Omega$ are easy to deal with, since one simply has to use
the appropriate $g_j$ and $\Omega_j$ 
values in Eqs. (\ref{i5}) and (\ref{ii9}).

Equation (\ref{ma0}) reveals a two-fold role of the el-ph interaction.
If the solution was just $G(i,j,\omega) = G_{\rm
d}(i,j,\tilde{\omega})$, it would mean that the renormalized
quasiparticle -- the polaron -- interacts with the bare impurity potential
$\hat{U}_0$. However, the second term shows that the impurity
potential is renormalized as well
\begin{equation}
\EqLabel{zao} \hat{U}_0 \rightarrow \hat{U}_0+\sum_{j}^{}
v_0(j,\omega) c^\dagger_jc_j,
\end{equation}
and is no longer local  and has significant
retardation effects through its $\omega$-dependence. In other words,
because of el-ph interactions, the dressed polaron interacts with a
renormalized, retarded disorder potential. 

Since $\hat{U}_0$ is treated exactly, we expect the validity of this
approximation to mirror that of the MA$^{(0)}$ for pure Holstein
($U=0$)  therefore to  worsen
in lower dimensions, when $g\sim t$ (such that the effective 1D
el-ph coupling $\lambda= g^2/(2t\Omega)\sim 1$; remember that the
approximation becomes exact for both $\lambda\rightarrow 0, \infty$) and for smaller
$\Omega$ \cite{MB2}. Even then, a larger $U$ improves the
accuracy as it pushes the GS to lower energies. However, for
small $U$ and $\Omega$ and for $\lambda \sim 1$ we need to go to a higher
level of the approximation.  Like for the homogeneous MA$^{(1)}$
solution \cite{MB2}, in IMA$^{(1)}$ we neglect the exponentially small
contributions only if there are $n\ge 2$ phonons present. The IMA$^{(1)}$
solution is similar to Eq. (\ref{ma0}), except $\tilde{\omega}$ is
now renormalized by the bulk $\Sigma_{\rm MA^{(1)}}(\omega) $
self-energy~\cite{MB2}, while the renormalized potential is also more
accurate:
$
v_1(j,\omega) = g^2x_{j,\omega}/\left[1-gx_{j,\omega}[A_2(j,\omega)-
A_1(j,\omega-\Omega)]\right]- \Sigma_{\rm MA^{(1)}}(\omega)
$ where $x_{j,\omega} = G_{\rm
  MA^{(0)}}(j,j,\omega-\Omega)$ is the MA$^{(0)}$ solution of
Eq. (\ref{ma0}) at a shifted energy. If necessary, one can go to a higher
IMA$^{(n)}$ ($n \ge 2$) level in the same way.

\begin{figure}[t]
\includegraphics[width=0.9\columnwidth]{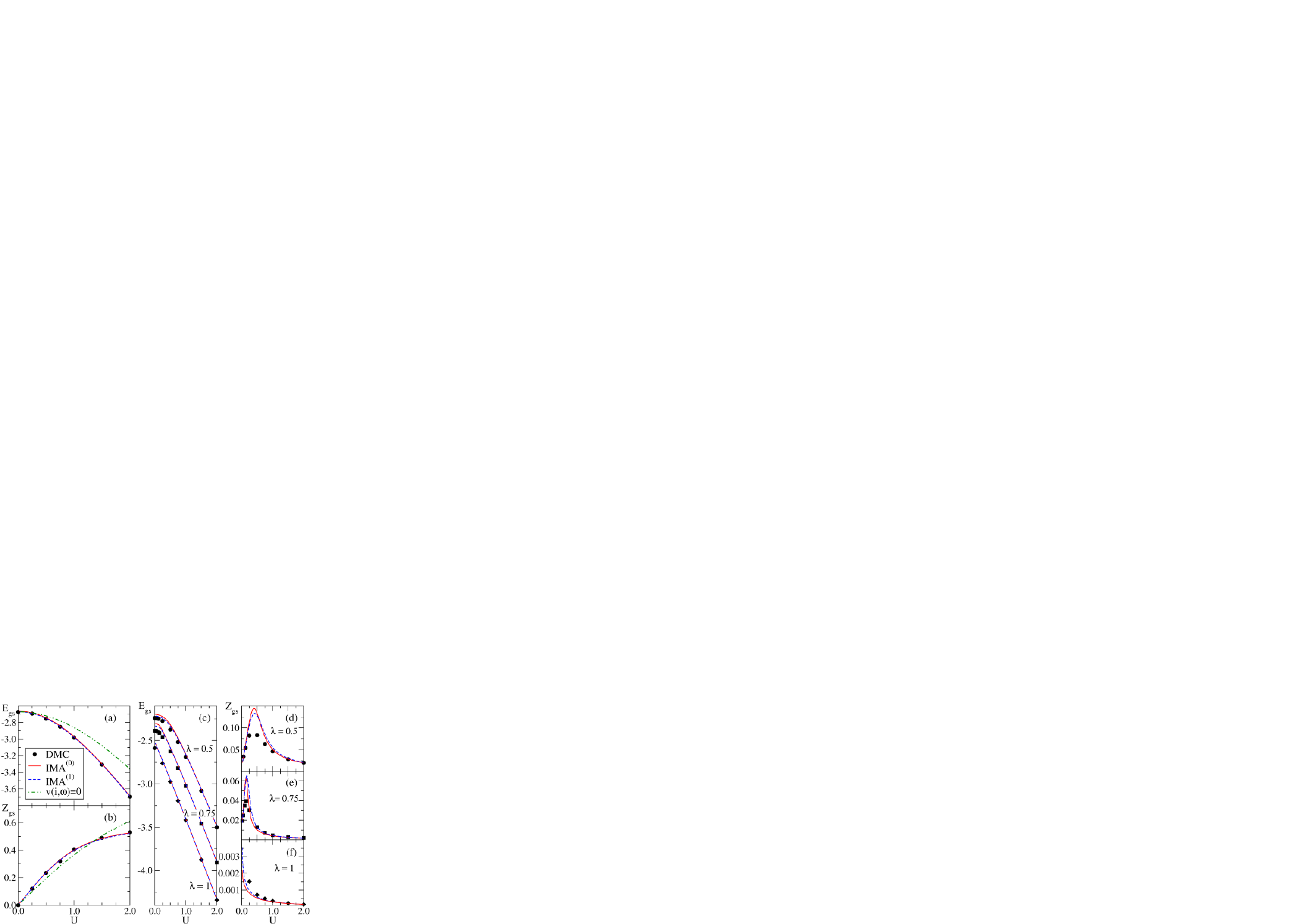}
\caption{(color online) (a) and (c) Ground-state energies; and (b),(d)-(f) Spectral weights at the
  impurity site vs. the impurity potential $U$, for $t=1$, $\Omega=2$
  and $g =1.5$ so that $\lambda = 0.5626$, in (a) and (b),
  respectively $\Omega=0.2$ and $g=\sqrt{0.2},   \sqrt{0.3},$
  $\sqrt{0.4}$ so that $\lambda=0.5, 0.75, 1$ in (c)-(f).}
\label{fig1}
\end{figure}
%
%
%

We gauge the accuracy of IMA for model (\ref{e1}) against DMC results.
Fig. \ref{fig1} shows the GS energy and quasiparticle weight $Z_{\rm
gs}$ at the impurity site vs. the impurity potential $U$.  The
agreement is very good in (a) and (b) even though these are 1D
results, where IMA is least accurate. This is partially due to the
large $\Omega=2t$ used~\cite{MB2}. For a worst-case scenario, we plot
in (c)-(f) results for a much smaller $\Omega/t =0.2$, for weak,
medium and strong couplings. Now we see clear differences, although
they are quantitative, not qualitative. For $E_{\rm gs}$, the
disagreements at small $U$ just mirrors the errors of MA for the
Holstein model ~\cite{MB2}. As expected, as $U$ increases and $E_{\rm
gs}$ moves to lower energies, the accuracy improves.  $Z_{\rm gs}$
shows more significant errors at small $U$, with IMA overestimating
the correct answer. This is not surprising, since for such small
$\Omega$ one expects many phonons to be created at many sites and a
higher level $n$ of IMA is needed. However, even levels $n=0,1$
capture the physics quite well.  Moreover, given the spectral weight
sum rules satisfied exactly (6 for IMA$^{(0)}$, 8 for IMA$^{(1)}$, see
Ref.~\cite{MB2}), we expect the spectral weight at all energies to be
similarly accurate.

In Fig. \ref{fig1} (a), (b) we also show what happens if we set the
additional potential $v_0(j,\omega) \rightarrow0$, i.e. we
use an ``instantaneous''
approximation (see e.g. Refs.~\cite{Bur08},\cite{Mac04}) where the 
el-ph coupling is assumed to renormalize the 
potential $\hat{U}_0 \rightarrow
\hat{U}_0-{g^2\over \Omega}\sum_{j}^{}
c^\dagger_jc_j$. This overall shift is given, in IMA, by the bulk
self-energy $\Sigma_{\rm MA^{(0)}}(\omega\approx E_{\rm
  gs})\approx -g^2/\Omega$  through the renormalized
$\tilde{\omega}$. Clearly, this works quite poorly even for this
rather small $\lambda$, and becomes considerably worse as $g$ and
therefore $v_0(j,\omega)$ increase.

\begin{figure}[t]
\includegraphics[width=0.98\columnwidth]{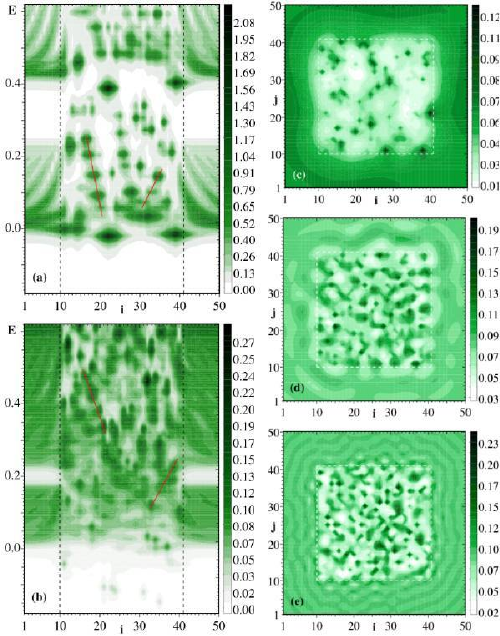}
\caption{(color online) (a) 1D LDOS $A(i,E)$. The polaron band and part of the second
bound state band are shown.  The vertical dashed lines mark the edges
of the disordered region. The slanted lines show the dispersion of
some features in the spectrum; (b) Analog of (a) for a 2D sample.  The
plot shows $A(i,j=25,\omega)$ vs $i$; (c,d,e) show the 2D $A(i,j,E)$
vs. $i,j$ for $E = 0, 0.25$ and $0.5$ respectively. In all plots the
energy is measured from the ground-state polaron energy of the
corresponding uniform system with $\lambda=1$, $\Omega=0.5t$.}
\label{fig2}
\end{figure}

In Fig. \ref{fig2} we show IMA results for a nontrivial problem whose
treatment by the DMC method is not feasible due to enormous
computational costs. Here we study the site and energy dependent map
of the LDOS of an area with randomly chosen el-ph effective coupling
$\lambda_i = g_i^2/(2t\Omega) \in [0.9, 1.1]$ embedded in an otherwise
uniform infinite system with 
$\lambda =1$. We do not add an on-site disorder potential, although it can
be included exactly as discussed above. The LDOS maps in
Fig.~\ref{fig2}(a),(b) show clear 
evidence of the nonlocal response of the system to
such inhomogeneities in the coupling constant, with various features
changing their position in space as the energy is changed and
$v_0(j,\omega)$ varies (the slanted lines
show some examples). This is more pronounced in 1D, since in 2D some
of these shifts proceed in the direction perpendicular to the line with $j=25$.
The retarded nature is apparent not only through the different-looking
LDOS maps at different energies inside the disordered region, but also
by the slow convergence towards the uniform bulk value and the
Friedel-like oscillations seen in the uniform region surrounding
it. As expected, the wavelength of these oscillations decreases with
increasing energy. 

\begin{figure}[t]
\includegraphics[width=0.6\columnwidth]{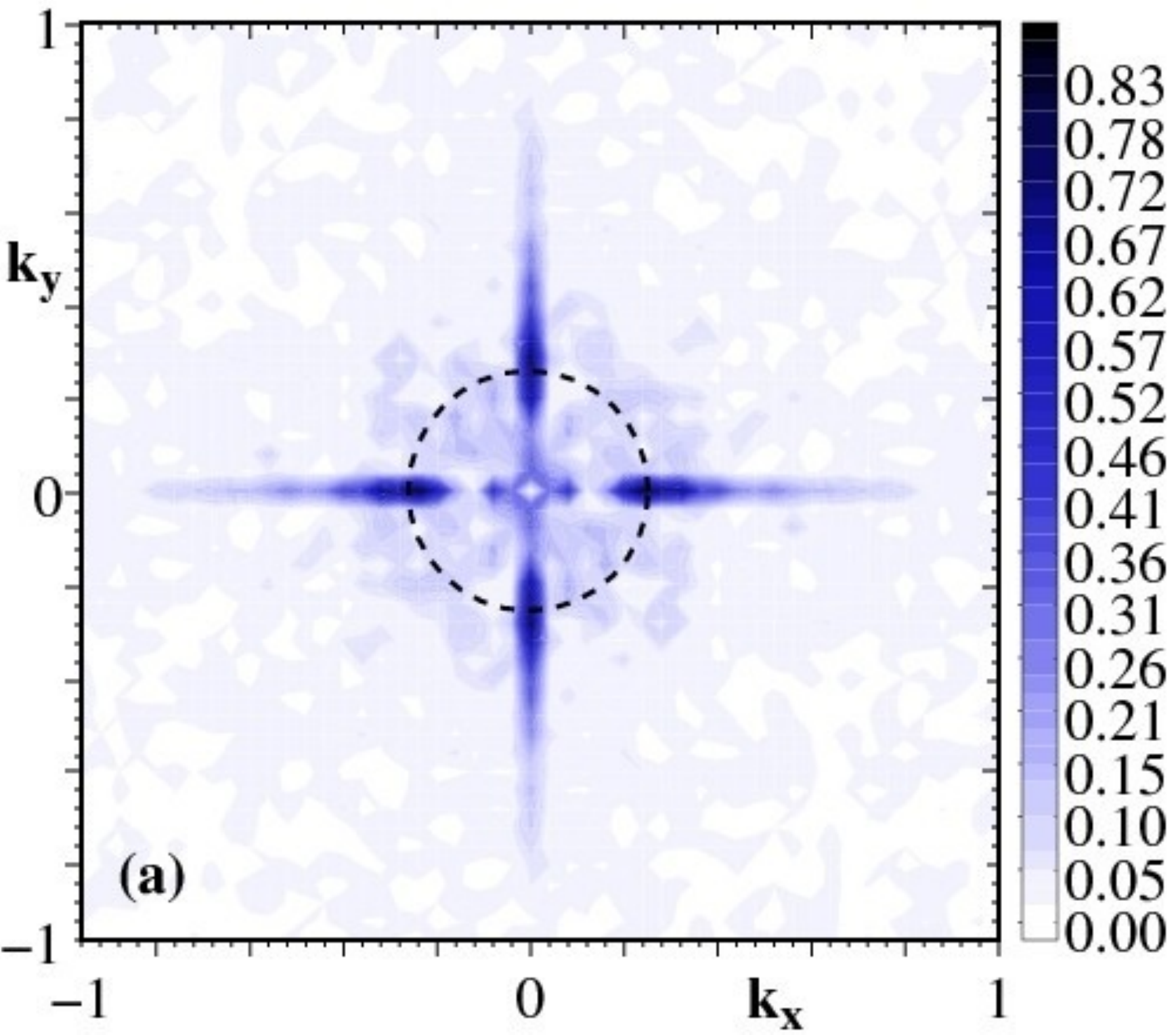}
\includegraphics[width=0.37\columnwidth]{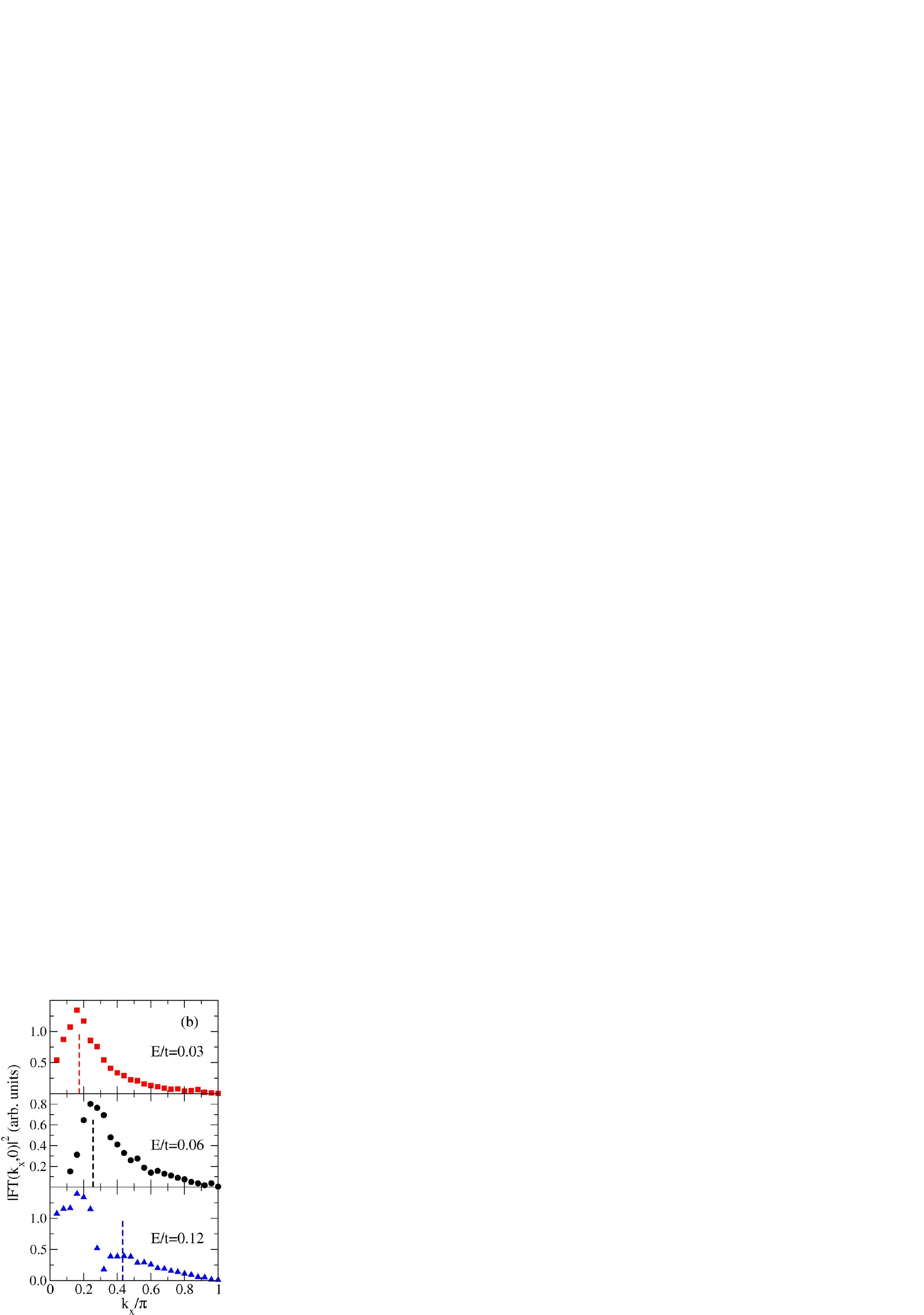}
\caption{(color online) (a) Fourier transform of the 2D LDOS map at
  energy $E=0.06t$ above the bulk ground-state;  (b) Same
  for $k_y=0$ and $E/t=0.03, 0.06, 0.12$. The
  lines show predicted scattering $\kv_s$ vectors. For more
  details, see text.}
\label{fig3}
\end{figure}

The 2D LDOS maps at different energies show no correlations, but
 their Fourier transforms (FT) should show peaks at $\kv$ values where
 scattering between {\em qp} states of that energy is likeliest. Such
 analysis is well known for STM data in cuprates, for scattering both
 on impurities~\cite{bis1} and inhomogeneities in the el-ph
 coupling~\cite{bis2}. In Fig. \ref{fig3}(a) we show the FT at
 $E=0.06t$ above the bulk polaron ground-state (we averaged over 
 FT of 200 LDOS maps such as shown in Fig. \ref{fig2}, but for 50x50 site
 inhomogeneous regions. We removed the $\kv =0$ peak for clarity). At
 such low energies, the bulk polaron dispersion $E(\kv) = -2t^*(\cos
 k_x + \cos k_y-2)\approx \hbar^2 \kv^2/(2m^*)$, so we expect signal
 up to a $k_s = 2 \sqrt{2m^*E/\hbar^2}$ for $\kv \rightarrow -\kv$
 scattering.  The 
 effective polaron mass is  $m^*/m=1.86$ for $\lambda=1,
 \Omega=0.5t$. We show $k_s$ 
 as a dashed line found to be in agreement 
 with the  FT data, as confirmed in Fig. \ref{fig3}(b)
 for several energies. This demonstrates that it is the polaron and
 not the bare particle that is scattered by 
inhomogeneities. We believe that the second  peak visible for
 higher energies is due to inelastic scattering between the first and
 second polaron bound states, but this needs further study. In any
 event, these results show agreement with the general phenomenology
 seen in the experimental data
 \cite{bis2,bis1} well
 beyond the single site disorder case usually considered
 theoretically~\cite{marcel}. 

Because IMA is very fast (2D LDOS maps such as shown in
Fig. \ref{fig2}(c) take about 50s to generate on an ordinary desktop;
moreover this type of calculation is ideal to parallelize, with
different CPUs for different energies) one can easily study very large
disordered regions, for different types of disorder in the on-site
potential (whether short- or long-range) and/or el-ph coupling or
phonon frequency, in any dimension. Such studies will reveal the
effects of each kind of disorder and the way in which 
multiple types of inhomogeneities do or do not
``interfere''. Moreover, such studies can be extended to other bare
$qp$ dispersions as well as models beyond
Holstein, such as systems with multiple phonon modes or with
electron-phonon coupling dependent on the phonon momentum, where the
bulk MA solutions are already known~\cite{MB5}.

In conclusion, we have studied the Holstein polaron problem with
disorder by developing an accurate, controllable and fast
computational method suitable for a
huge class of problems that were unaccessible until now.

{\em Acknowledgments:} NSERC and CIFAR
(M.B.),  RFBR 07-02-0067a (A.S.M.), and  Grant-in-Aids No. 15104006,
No. 16076205, No. 17105002, No. 19048015, and NAREGI Japan (N.N.).

\end{document}